\documentclass[final,5p,times,twocolumn]{elsarticle}
\usepackage{amsmath, amssymb}
\usepackage{graphicx}
\usepackage{booktabs}
\usepackage{tabularx}
\usepackage{array}
\usepackage{placeins}
\usepackage{float}
\usepackage{caption}
\usepackage[hidelinks]{hyperref}
\biboptions{numbers,sort&compress}
\journal{Applied Energy}

\begin{document}

\begin{frontmatter}
\title{Quantifying the hosting capacity value of workload flexibility for AI data centers: A two-stage electricity–computation co-optimization framework}
\author[edinburgh]{Jiakai Wu}
\ead{s2902684@ed.ac.uk}
\author[edinburgh]{Pengyu Ren}
\ead{s2192464@ed.ac.uk}
\author[edinburgh]{Yifan Wang}
\ead{s2154060@ed.ac.uk}
\author[edinburgh]{Wei Sun\corref{cor1}}
\ead{w.sun@ed.ac.uk}
\cortext[cor1]{Corresponding author.}
\address[edinburgh]{School of Engineering, The University of Edinburgh,
Edinburgh EH9 3FB, Scotland, United Kingdom}

\begin{abstract}
Grid connection constraints increasingly limit the expansion of AI-era data centers, yet the additional hosting capacity enabled by computational workload flexibility remains insufficiently quantified. This paper develops a two-stage electricity--computation co-optimization framework to quantify the individual and coordinated hosting capacity values of spatial and temporal workload flexibility. Data center capacities are determined endogenously, with computational task arrivals scaling with admitted capacity. Spatial routing of eligible interactive requests and temporal deferral of batch workloads are linked to server operation and facility power demand, subject to workload completion requirements and transmission constraints. The first stage maximizes aggregate hosting capacity; the second fixes the resulting site capacities and optimizes annual workload scheduling and generation dispatch to minimize generation and carbon costs. An 8760-hour case study with three data centers on a modified IEEE 118-bus system shows that temporal and spatial flexibility individually increase hosting capacity by 21.26\% and 5.37\%, respectively. Their coordinated use raises hosting capacity from 4635.1~MW to 6231.5~MW, a 34.44\% increase without network reinforcement. The coordinated gain exceeds the sum of the individual gains by 361.9~MW, demonstrating complementarity between spatial migration and temporal deferral. By redistributing computational demand across locations and operating hours, these mechanisms enable fuller utilization of existing transmission capacity. The findings establish workload flexibility as a quantifiable planning resource for accommodating additional data center capacity within existing transmission networks.
\end{abstract}

\end{frontmatter}

\begin{table*}[t]
\centering
\begin{minipage}{0.98\textwidth}
\hrule
\vspace{6pt}
\textbf{Nomenclature}
\vspace{4pt}

\footnotesize
\setlength{\tabcolsep}{1pt}
\renewcommand{\arraystretch}{1.04}

\begin{minipage}[t]{0.485\textwidth}
\textit{A. Abbreviations and sets}

\begin{tabularx}{\linewidth}{@{}>{\raggedright\arraybackslash}p{0.18\linewidth}X@{}}
DC & Data center. \\
ECC & Electricity--computation co-optimization. \\
HC & Hosting capacity. \\
IT & Information technology. \\
PUE & Power usage effectiveness. \\
\(\mathcal N,b\) & Set and index of transmission-network buses. \\
\(\mathcal L,\ell\) & Set and index of transmission lines. \\
\(\mathcal G,g\) & Set and index of conventional generators. \\
\(\mathcal R,r\) & Set and index of renewable generators. \\
\(\mathcal D,n,n'\) & Set and indices of data centers. \\
\(\mathcal T,t,t'\) & Set and indices of operating periods. \\
\(\mathcal C,c\) & Set and index of flexibility cases. \\
\(i(n)\) & Connection bus of DC \(n\). \\
\(o(\ell),e(\ell)\) & Sending and receiving buses of line \(\ell\). \\
\end{tabularx}

\vspace{4pt}
\textit{B. Parameters}

\begin{tabularx}{\linewidth}{@{}>{\raggedright\arraybackslash}p{0.21\linewidth}X@{}}
\(T\) & Number of periods in the annual horizon. \\
\(T^{\mathrm{D}}\) & Maximum batch workload delay. \\
\(\mu_n\) & Processing rate of a server at DC \(n\). \\
\(\bar U\) & Maximum server utilization. \\
\(\rho_n\) & Installed server density (servers/MW). \\
\(p^{\mathrm{idle}}_n\) & Idle server power consumption at DC \(n\) (kW). \\
\(p^{\mathrm{peak}}_n\) & Peak server power consumption at DC \(n\) (kW). \\
\(\eta^{\mathrm{DC}}\) & DC power usage effectiveness (PUE). \\
\(p^{\mathrm{load}}_{b,t}\) & Conventional load at bus \(b\) in period \(t\) (MW). \\
\(\bar p^{\mathrm R}_{r,t}\) & Available power from renewable generator \(r\) in period \(t\) (MW). \\
\(\underline p^{\mathrm G}_g,\bar p^{\mathrm G}_g\) & Power output limits of conventional generator (MW). \\
\(c_{2,g},c_{1,g},c_{0,g}\) & Quadratic generation cost coefficients of generator \(g\). \\
\(\gamma_g\) & Generator emission factor of generator \(g\) (tCO\(_2\)/MWh). \\
\(c^{\mathrm{CO2}}\) & Carbon price (£/tCO\(_2\)). \\
\(x_\ell,\bar f_\ell\) & Line reactance and active-power rating of line \(\ell\). \\
\end{tabularx}
\end{minipage}%
\hfill
\begin{minipage}[t]{0.485\textwidth}

\begin{tabularx}{\linewidth}{@{}>{\raggedright\arraybackslash}p{0.16\linewidth}X@{}}
\(S^{\mathrm{base}}\) & Network base power (MVA). \\
\(b^{\mathrm{ref}}\) & Reference bus. \\
\end{tabularx}

\vspace{4pt}
\textit{C. Decision variables}

\begin{tabularx}{\linewidth}{@{}>{\raggedright\arraybackslash}p{0.14\linewidth}X@{}}
\(H_{n,c}\) & Hosting capacity at DC \(n\) in case \(c\) (MW). \\
\(H_{n,c}^\star\) & Optimized hosting capacity at DC \(n\) in case \(c\) (MW). \\
\(H_c^\star\) & Aggregate optimized hosting capacity in case \(c\) (MW). \\
\(\mathcal V_c^{1},\mathcal V_c^{2}\) & Decision variable sets for Stages 1 and 2. \\
\(w^{\mathrm 0}_{n,t}\) & Total workload arrival. \\
\(w^{\mathrm I0}_{n,t}\) & Interactive workload arrival at DC \(n\) at time \(t\). \\
\(w^{\mathrm B0}_{n,t}\) & Batch workload arrival at DC \(n\) at time \(t\). \\
\(w^{\mathrm{Inflex}0}_{n,t}\) & Inflexible workload arrival at DC \(n\) at time \(t\). \\
\(w^{\mathrm{Inflex}}_{n,t}\) & Processed inflexible workload at DC \(n\) at time \(t\). \\
\(w^{\mathrm I}_{n,t}\) & Processed interactive workload at DC \(n\) at time \(t\). \\
\(w^{\mathrm B}_{n,t}\) & Processed batch workload at DC \(n\) at time \(t\). \\
\(w_{n,t}\) & Total processed workload at DC \(n\) at time \(t\). \\
\(w^{\mathrm I}_{n,n',t}\) & Interactive workload shifted \(n\to n'\) at time \(t\). \\
\(w^{\mathrm B}_{n,t,t'}\) & Batch workload deferred \(t\to t'\) at DC \(n\). \\
\(m_{n,t}\) & Active server equivalent. \\
\(u_{n,t}\) & Average server utilization. \\
\(p^{\mathrm{IT}}_{n,t}\) &  IT power consumption at DC \(n\) in period \(t\) (MW). \\
\(p^{\mathrm{nonIT}}_{n,t}\) & Aggregated non-IT power demand at DC \(n\) (MW). \\
\(p^{\mathrm{DC}}_{n,t}\) & Power consumption at DC \(n\) at time \(t\) (MW). \\
\(p^{\mathrm G}_{g,t}\) & Power output of generator \(g\) at time \(t\) (MW). \\
\(p^{\mathrm R}_{r,t}\) & Power output of renewable generator \(r\) at time \(t\) (MW). \\
\(p^{\mathrm{curt}}_{r,t}\) & Renewable curtailment of renewable generator \(r\) at time \(t\) (MW). \\
\(\theta_{b,t}\) & Voltage angle at bus \(b\) at time \(t\). \\
\(f_{\ell,t}\) & Active power flow on line \(\ell\) at time \(t\) (MW). \\
\(C^{\mathrm{oper}}\) & Total system operating cost. \\
\(C^{\mathrm G}\) & Conventional generation cost. \\
\(C^{\mathrm{carbon}}\) & Carbon emission cost. \\
\(E^{\mathrm{carbon}}\) & Generator emissions. \\
\(E^{\mathrm{curt}}\) & Renewable energy curtailment. \\
\end{tabularx}
\end{minipage}
\vspace{5pt}
\hrule
\end{minipage}
\end{table*}

\section{Introduction}

\subsection{Background and motivation}

Driven by artificial intelligence (AI) and continued digitization, data centers (DCs) are emerging as one of the fastest-growing and most concentrated sources of electricity demand. Global DC electricity consumption reached approximately 415~TWh in 2024 and is projected to more than double to around 945~TWh by 2030, with AI expected to drive the largest share of this increase
\cite{IEA2025EnergyAI}. In the United States, DCs accounted for approximately 4.4\% of national electricity in 2023, and this share is projected to reach 6.7--12\% by 2028 \cite{Shehabi2024DataCenterEnergy}. At the project level, new hyper-scale facilities are being developed with connection capacities of 100-1000~MW, and utilities are receiving individual connection requests of 1~GW or more
\cite{EPRI2024PoweringDataCenters}. NERC further reports that emerging large
loads range from multi-megawatt facilities to multi-gigawatt complexes requiring
transmission level connections and associated network upgrades
\cite{NERC2025EmergingLargeLoads}. These developments concentrate substantial
demand at individual points of interconnection and within a limited number of
regional clusters. Consequently, the local network impact of an AI-era DC can
be considerably greater than its contribution to national annual electricity
consumption. Although improvements in hardware efficiency and server utilization
have moderated historical growth in DC electricity consumption
\cite{Masanet2020Recalibrating}, they do not reduce the need to supply
increasingly large blocks of connection capacity at specific network locations.

The rapid development cycle of DCs further intensifies this locational challenge. New hyper-scale capacity can potentially be developed within one to two years, while a typical DC can become operational within two to three years \cite{EPRI2024PoweringDataCenters}. By comparison, new transmission infrastructure typically requires four to eight years to develop in advanced economies, while lead times for critical equipment such as transformers and cables have also lengthened \cite{IEA2025EnergyAI}. In established European DC hubs, grid connection times can extend to ten years because of network congestion and the need for substantial grid reinforcement \cite{Ember2025GridsDataCentres}. This mismatch means that a DC project may be ready for operation well before the fundamental electricity infrastructure required for its requested connection. Consequently, grid constraints could delay approximately 20\% of the global DC capacity planned for deployment by 2030 \cite{IEA2025EnergyAI}. The relevant planning question is whether the existing network can accommodate the proposed DC capacity at the requested connection point and target date. For an existing network, this capacity is further constrained by the transmission headroom available across the operating hours in which DC demand coincides with other system demand. DC hosting capacity is thus inherently dependent on both spatial connection location and temporal operational conditions.

Under a conventional firm connection, the demand profile associated with the proposed DC connection capacity must remain feasible with respect to network operating constraints across all operating hours. The resulting HC may therefore be constrained by a small number of congested node-hour combinations, even when transmission headroom remains available at other locations or times. Workload flexibility provides a means of coordinating DC electricity demand with these spatially and temporally varying network conditions. Previous research has demonstrated that dispatchable computing loads can improve grid utilization by relocating computational demand \cite{Kim2017DispatchableLoads}. For DCs, this coordination begins with the computational service: facility electricity demand is determined by the workloads processed through IT equipment and the power required by cooling and other supporting systems  \cite{Huang2020DataCenterProsumers}. Latency-sensitive interactive requests can be routed among geographically distributed DCs, whereas delay-tolerant batch workloads can be deferred within prescribed completion windows. These scheduling decisions change server utilization and facility power demand while preserving the associated computing
service. Fleet-scale implementations have further demonstrated the practical feasibility of temporal workload scheduling across geographically distributed computing resources \cite{Radovanovic2023CarbonAware}. This coupling motivates an
electricity-computation co-optimization framework in which workload scheduling,
DC power consumption, power system operation, and transmission feasibility are
represented consistently. Within this framework, the value of DC flexibility in hosting capacity is quantified as the additional network-constrained HC enabled relative to an inflexible DC workload configuration.

\subsection{Literature review and research gap}

Research on DC flexibility has evolved along three connected directions:
modeling the computational sources of DC flexible power demand, coordinating
that demand with power system operation, and assessing DC siting and
hosting capacity. Table~\ref{tab:literature_comparison} compares the
studies most directly related to network-constrained DC hosting capacity assessment. The comparison distinguishes the representation of DC power demand, the available workload flexibility dimensions, the scale and configuration of the DC portfolio, the modeled network constraints, and the treatment of connection
capacity. The final column identifies whether flexibility is valued through the
additional HC enabled by DC workload scheduling.

\begin{table*}[t]
\caption{Review and comparison of relevant literature.}
\label{tab:literature_comparison}
\centering
\scriptsize
\setlength{\tabcolsep}{2.4pt}
\renewcommand{\arraystretch}{1.12}
\begin{tabular*}{\textwidth}{@{\extracolsep{\fill}}l l c c c c c c
    >{\centering\arraybackslash}p{2.15cm}@{}}
\toprule
Ref. & DC demand model
& Workload flexibility
& \shortstack{DC scale}
& \shortstack{DC configuration}
& \shortstack{Network level}
& \shortstack{Network constraints}
& \shortstack{Capacity formulation}
& \shortstack{Flexibility value in \\hosting capacity} \\
\midrule
\cite{Chen2021OperationalFlexibility} & \mbox{Workload to power} & SF\&TF
& $\leq 100$~MW & Multiple & DN & $\checkmark$ & -- & -- \\
\cite{Cao2022DataDrivenFlexibility} & Trace-derived power & TF
& $\leq 100$~MW & Single & -- & -- & -- & -- \\
\cite{ZhangZavala2022Remunerating} & Virtual-link load & SF\&TF
& $\leq 100$~MW & Multiple & TN & $\checkmark$ & -- & -- \\
\cite{Li2024ComputationPower} & \mbox{Workload-to-power} & SF
& $\leq 100$~MW & Multiple & DN & $\checkmark$ & -- & -- \\
\cite{Zeng2024BilevelPlanning} & \mbox{Workload-to-power} & SF\&TF
& $\leq 100$~MW & Multiple & TN & $\checkmark$ & EN & -- \\
\cite{Shao2025StochasticPlanning} & \mbox{Workload-to-power} & SF
& GW-scale & Multiple & TN & $\checkmark$ & EN & -- \\
\cite{Ye2026Harvesting} & \mbox{Workload-to-power} & SF\&TF
& $\leq 100$~MW & Multiple & TN & $\checkmark$ & -- & -- \\
\cite{Kim2026FlexibilityAware} & Flexible-load envelope & TF
& GW-scale & Single & TN & $\checkmark$ & EX & $\checkmark$ \\
\cite{FangLi2026DCHostingCapacity} & Static power load & --
& $\leq 100$~MW & Single & DN & $\checkmark$ & EN+EX & -- \\
\cite{Manassero2026ResilienceAware} & Static power load & --
& GW-scale & Multiple & TN & $\checkmark$ & EX & -- \\
\cite{Samir2026AIHostingPEM} & GPU power profile & --
& GW-scale & Multiple & TN & $\checkmark$ & EN+EX & -- \\
\textbf{This paper} & \textbf{\mbox{Workload-to-power}} & \textbf{SF\&TF}
& \textbf{GW-scale} & \textbf{Multiple} & \textbf{TN}
& $\boldsymbol{\checkmark}$ & \textbf{EN+EX}
& $\boldsymbol{\checkmark}$ \\
\bottomrule
\end{tabular*}
\vspace{2pt}
\begin{minipage}{\textwidth}
\footnotesize
SF: spatial workload flexibility; TF: temporal workload flexibility;
DN: distribution network; TN: transmission network; EN: endogenous DC hosting
capacity; EX: explicit hosting capacity; EN+EX: endogenous DC capacity within an explicit hosting capacity.

\end{minipage}
\end{table*}

The first research direction establishes the computational basis of flexible DC
power demand. Workload-to-power formulations translate the allocation and
execution of computational tasks into server activation, utilization, and
facility power consumption, enabling spatial and temporal scheduling to
be represented within power system operation
\cite{Chen2021OperationalFlexibility,Li2024ComputationPower,Ye2026Harvesting}.
Data-driven methods complement these formulations by estimating the deferrable
power associated with periodic batch workloads from operational traces
\cite{Cao2022DataDrivenFlexibility}. The amount of available flexibility depends on the service characteristics of the underlying workloads. QoS-aware demand response studies show that job scheduling and server level power control can provide grid response while maintaining computing service
requirements \cite{Zhang2022HPCDemandResponse}. Trace-driven analysis across
interactive and batch workloads further demonstrates that the available
temporal and spatial shifting depend on job duration, completion deadlines,
service level objectives, and the available destination regions
\cite{Sukprasert2024Limitations}. Workload flexibility is therefore a service-constrained capability defined by the extent to which computation can be shifted across locations and time while satisfying its service requirements.

The second direction incorporates this capability into power system
coordination. Virtual-link formulations represent temporal deferral and
spatial migration as non-physical transfer pathways in network-constrained
market clearing, allowing their effects on nodal prices and system operation to
be evaluated \cite{ZhangZavala2022Remunerating}. Related incentive mechanisms
coordinate the spatial-temporal response of DCs with power system requirements
\cite{Wu2023SpatiotemporalFlexibility}, while detailed
electricity-computation models schedule heterogeneous workloads under network and uncertainty constraints \cite{Ye2026Harvesting}. These capabilities have been applied to renewable energy matching and carbon-aware operation \cite{Yang2023CarbonManagement,Riepin2025CleanComputing},
transmission and server planning \cite{Wan2023NetworkTransition}, balancing and
non-wire alternatives \cite{Cao2024NonWireAlternative}, post-fault restoration
\cite{Jian2024SupplyRestoration}, and distribution network congestion management
\cite{Zhang2025ProactiveWorkload}. Collectively, these studies demonstrate how
workload scheduling changes the location and time of DC power demand and
produces measurable operational, economic, and environmental outcomes. Their
assessments generally begin from a prescribed DC portfolio or determine DC
capacity through cost, carbon, or investment objectives; the maximum hosting
capacity supported by workload flexibility is not their principal assessment
quantity.

The third direction moves from system operation to DC siting, sizing, and
interconnection capacity. Joint planning models make DC location and capacity
endogenous alongside electricity-system investment and operation
\cite{Zeng2024BilevelPlanning,Shao2025StochasticPlanning}.
Flexibility-aware connection screening evaluates whether GW-scale projects can
be accommodated at candidate locations under firm, pause, and shift
envelopes \cite{Kim2026FlexibilityAware}. Explicit HC formulations have also
quantified the admissible DC loads supported by batteries, dispatchable
generation, and reactive power resources in distribution networks
\cite{FangLi2026DCHostingCapacity}. At the transmission level, recent studies
assess feasible allocations of multiple DCs under contingency requirements
\cite{Manassero2026ResilienceAware} and examine how electrolyzers can support
AI-training DCs subject to voltage and frequency constraints
\cite{Samir2026AIHostingPEM}. These studies bring DC power demand directly into
network-capacity assessment. Flexible demand is typically represented through prescribed DC power envelopes or load profiles, or provided by additional flexibility resources, rather than being endogenously derived from computational workloads whose processing is scheduled across locations and time.

Taken together, the existing literature leaves the relationship between workload flexibility and network hosting capacity insufficiently resolved. A hosting capacity-centered assessment should simultaneously make the capacities of multiple DCs endogenous and preserve workload completion under spatial and temporal scheduling under transmission network constraints. It must also evaluate comparable flexibility configurations on the same network so that differences in hosting capacity can be attributed directly to workload flexibility. Such a formulation makes it possible to quantify not only the additional DC capacity enabled by flexibility, but also the workload adjustments and network operating conditions through which these hosting capacity gains are realized.

\subsection{Research questions and contributions}

The literature review identifies two complementary research strands. Electricity-computation coordination studies have established how spatial and temporal workload scheduling can improve the operation of a DC portfolio, while hosting capacity studies have quantified the connection capacity of electricity networks under static or exogenously prescribed DC power demand. Connecting these two strands requires a framework that makes multiple DC hosting capacities endogenous, derives their power demand from computational workloads, and quantifies workload flexibility through the additional hosting capacity it enables. This forms the central focus of the present study. Accordingly, this paper addresses three research questions. First, how much aggregate DC hosting capacity can a transmission network accommodate? Second, how much additional hosting capacity is enabled by spatial and temporal flexibility individually, and how does their coordination change these gains? Third, how is the DC flexibility activated across locations and time, and how does this activation affect network operation to support additional hosting capacity? To answer these questions, the main contributions are as follows:

\begin{enumerate}
    \item An electricity-computation co-optimization model is developed for assessing the hosting capacity of multiple DCs connected to a transmission network. The hosting capacity of each DC determines its computational task arrivals, which are mapped through workload scheduling, server operation, and DC power demand seen by the network.

    \item A two-stage optimization framework is formulated that connects the planning value of workload flexibility with network operation. The framework first determines the maximum aggregate DC hosting capacity, and subsequently optimizes annual workload scheduling and power system operation at the resulting capacity allocation. This provides a unified assessment of how much DC hosting capacity is enabled by flexibility and how these gains are realized through coordinated workload scheduling and power system operation.

    \item The mechanism through which DC workload flexibility enables additional hosting capacity is characterized. Spatial migration and temporal deferral align DC power demand with available network headroom across locations and time. By separating the individual contributions of spatial and temporal flexibility from the interaction effect, the analysis reveals how workload scheduling converts available network headroom into DC hosting capacity.

\end{enumerate}

These contributions extend ECC from the operational use of a given
DC portfolio to the planning question of how much capacity a fixed network can
credibly host. They also provide a service-preserving basis for valuing
computational flexibility in flexible or non-firm connection assessments.

\subsection{Paper organization}

The remainder of this paper is organized as follows. Section~2 develops the DC
power, IT equipment, and workload management models. Section~3 formulates the
problem statement and develops the proposed two-stage hosting capacity
assessment framework. Section~4 defines the hosting capacity value metrics and
introduces the study system, experimental settings, comparative results,
flexibility activation mechanisms, and system impact analysis.
Section~5 summarizes the conclusions, planning implications, and
priorities for future work.

\section{Modeling of data center power and workload management}
\label{sec:modeling}

This section develops a model of DC power consumption and workload scheduling.
DC power consumption is jointly determined by the computational tasks processed
through IT equipment and the electricity required by cooling and other
facility-support systems. Among these components, IT equipment and cooling systems account for the majority of DC power consumption, with IT equipment representing the largest share \cite{VanGeet2024BestPractices}. Therefore, accurately modeling the power-consumption characteristics of computation is fundamental to constructing a comprehensive DC power model. The formulation
connects three interdependent layers: workload scheduling, server operation,
and DC power consumption. Workload arrivals scale with the DC hosting capacity
and are divided into interactive, batch, and inflexible components. The workload
processed at different locations and times determines active-server requirements
and utilization, which are mapped to IT power and subsequently to total DC power
demand. This workload-to-power model provides the DC-side foundation for the
methodology developed in Section~\ref{sec:methodology}.
\subsection{Data center power modeling}
\label{sec:dc_power_model}

The total power demand of a DC comprises IT power and the power required by cooling and other facility-support systems. These non-IT requirements are represented in aggregate using a fixed power usage effectiveness (PUE), denoted by \(\eta^{\mathrm{DC}}\) \cite{Shehabi2024DataCenterEnergy}. The relationship between IT power and total facility power is expressed as:

\begin{equation}
    p^{\mathrm{DC}}_{n,t}=p^{\mathrm{IT}}_{n,t}
    +p^{\mathrm{other}}_{n,t},
    \qquad
    p^{\mathrm{other}}_{n,t}
    =(\eta^{\mathrm{DC}}-1)p^{\mathrm{IT}}_{n,t}
    \label{eq:dc_power_pue}
\end{equation}

In Eq.~\eqref{eq:dc_power_pue}, \(p^{\mathrm{other}}_{n,t}\) represents the aggregated non-IT power demand, including cooling, lighting and other facility-support loads. Thus, changes in IT power caused by workload scheduling are translated into changes in the total DC power demand supplied by the network.

\subsection{IT equipment modeling}
\label{sec:it_equipment_model}

IT power consumption is modeled as a function of the number of active servers and their utilization \cite{Ding2022CoordinatedOperation}. Let \(m_{n,t}\) denote the continuous equivalent number of active servers at DC \(n\) in period \(t\), and let \(u_{n,t}\) denote their average utilization. The power consumption of an active server increases linearly from its idle power \(p^{\mathrm{idle}}_n\) to its fully utilized power \(p^{\mathrm{peak}}_n\). The aggregate IT power demand is therefore:
\begin{equation}
    p^{\mathrm{IT}}_{n,t}
    =m_{n,t}\left[p^{\mathrm{idle}}_n
    +u_{n,t}\left(p^{\mathrm{peak}}_n-p^{\mathrm{idle}}_n\right)\right]
    \label{eq:it_power_utilization}
\end{equation}
The first term accounts for the idle power of the active servers, while the second represents the additional power associated with their computational load. IT power therefore depends on both the number of servers activated and the workload assigned to them.

Server utilization is determined by the workload processed during the period relative to the processing capacity of the active servers. This relationship is:
\begin{equation}
    u_{n,t}=\frac{w_{n,t}}{\mu_n m_{n,t}},
    \qquad m_{n,t}>0,
    \label{eq:server_utilization}
\end{equation}
Here, \(w_{n,t}\) is the workload processing rate after spatial routing and temporal scheduling, and \(\mu_n\) is the processing rate of one fully utilized server. Accordingly, \(w_{n,t}\) refers to the computation executed at DC \(n\) in period \(t\), which may differ from the workload originally arriving there. Substituting Eq.~\eqref{eq:server_utilization} into Eq.~\eqref{eq:it_power_utilization} links the scheduled workload and active-server requirement to IT power.

\subsection{Workload modeling}
\label{sec:workload_model}

Workload flexibility is represented through three classes with different processing requirements. Interactive workloads can be routed among eligible DCs for processing within the same period, whereas batch workloads can be deferred within a prescribed completion window at their original DC. Inflexible workloads must be processed at their original location and time. 

In all cases, workload arrivals scale with admitted DC capacity, so that additional hosting capacity carries an additional computing service obligation. The workload arrival is expressed as:
\begin{equation}
    w^0_{n,t}=H_{n,c}\bar a_{n,t},
    \qquad n\in\mathcal D,\;t\in\mathcal T
    \label{eq:capacity_dependent_workload_arrivals}
\end{equation}
\begin{equation}
    w^0_{n,t}=w^{\mathrm I0}_{n,t}
    +w^{\mathrm B0}_{n,t}+w^{\mathrm{Inflex}0}_{n,t}
    \label{eq:workload_arrival_decomposition}
\end{equation}
The following formulations distinguish the workload originally arriving at each DC from the workload ultimately processed there and constrain how processing can be redistributed across locations and time.

\subsubsection{Interactive workloads}
\label{sec:interactive_workload}

Interactive workloads are subject to strict latency requirements. However, eligible task requests can be routed to other eligible DCs for processing, providing spatial flexibility.

Let
\(w^{\mathrm I}_{n,n',t}\) be the request volume shifted from DC
\(n\) to DC \(n'\neq n\) in period \(t\), and let
\(w^{\mathrm I}_{n,t}\) denote the interactive workload ultimately processed at
DC \(n\). The processed interactive workload is given by:
\begin{equation}
    \small
    w^{\mathrm I}_{n,t}=w^{\mathrm I0}_{n,t}
    +\sum_{n'\in\mathcal D_n}w^{\mathrm I}_{n',n,t}
    -\sum_{n'\in\mathcal D_n}w^{\mathrm I}_{n,n',t}
    \label{eq:interactive_routing_balance}
\end{equation}
\begin{equation}
    \sum_{n'\in\mathcal D_n}w^{\mathrm I}_{n,n',t}
    \leq w^{\mathrm I0}_{n,t}
    \label{eq:interactive_routing_limit}
\end{equation}

Equation \eqref{eq:interactive_routing_balance} explicitly captures the requests shifted into and out of each data center, whereas Eq.~\eqref{eq:interactive_routing_limit} ensures that the amount of workload shifted from a DC does not exceed the interactive workload originally arriving at that DC. Summing Eq.~\eqref{eq:interactive_routing_balance} across all data centers causes all transfers between DCs to cancel, thereby preserving the total volume of arriving requests in each hour.

\subsubsection{Batch workloads}
\label{sec:batch_workload}

Batch workloads comprise delay-tolerant computational tasks, including AI training and data-management tasks, that can be completed within prescribed time windows. Deferring eligible batch workloads within these windows allows DCs to redistribute computing demand across time, providing temporal flexibility.

Let the variable \(w^{\mathrm B}_{n,t,t'}\) denote the batch workload arriving in period \(t\) and deferred to period \(t'\). For the periods \(1\leq t\leq T^{\mathrm D}\), the workload balance and deferral limit are:
\begin{equation}
    \small
    w^{\mathrm B}_{n,t}=w^{\mathrm B0}_{n,t}
    +\sum_{\tau=1}^{t-1}w^{\mathrm B}_{n,\tau,t}
    -\sum_{\tau=t+1}^{t+T^{\mathrm D}}w^{\mathrm B}_{n,t,\tau}
    \label{eq:batch_initial_balance}
\end{equation}
\begin{equation}
    \sum_{\tau=t+1}^{t+T^{\mathrm D}}w^{\mathrm B}_{n,t,\tau}
    \leq w^{\mathrm B0}_{n,t}
    \label{eq:batch_initial_limit}
\end{equation}

For the periods \(T^{\mathrm D}+1\leq t\leq T-T^{\mathrm D}\), the workload balance and deferral limit are:
\begin{equation}
    \small
    w^{\mathrm B}_{n,t}=w^{\mathrm B0}_{n,t}
    +\sum_{\tau=t-T^{\mathrm D}}^{t-1}w^{\mathrm B}_{n,\tau,t}
    -\sum_{\tau=t+1}^{t+T^{\mathrm D}}w^{\mathrm B}_{n,t,\tau}
    \label{eq:batch_interior_balance}
\end{equation}
\begin{equation}
    \sum_{\tau=t+1}^{t+T^{\mathrm D}}w^{\mathrm B}_{n,t,\tau}
    \leq w^{\mathrm B0}_{n,t}
    \label{eq:batch_interior_limit}
\end{equation}

For the terminal periods \(T-T^{\mathrm D}+1\leq t\leq T\), the scheduling window is truncated at the end of the modeled year:
\begin{equation}
    \small
    w^{\mathrm B}_{n,t}=w^{\mathrm B0}_{n,t}
    +\sum_{\tau=t-T^{\mathrm D}}^{t-1}w^{\mathrm B}_{n,\tau,t}
    -\sum_{\tau=t+1}^{T}w^{\mathrm B}_{n,t,\tau}
    \label{eq:batch_terminal_balance}
\end{equation}
\begin{equation}
    \sum_{\tau=t+1}^{T}w^{\mathrm B}_{n,t,\tau}
    \leq w^{\mathrm B0}_{n,t}
    \label{eq:batch_terminal_limit}
\end{equation}
\begin{equation}
    \sum_{t=1}^{T}w^{\mathrm B}_{n,t}
    =\sum_{t=1}^{T}w^{\mathrm B0}_{n,t}
    \label{eq:batch_annual_completion}
\end{equation}
Considering the temporal flexibility of workload and its scheduling cycle, the three temporal domains are defined to be mutually disjoint and collectively span the complete 8760-hour horizon.

\subsubsection{Total workloads}
\label{sec:total_workload}

Inflexible workloads are processed at their original DC in their arrival period.
\begin{equation}
    w^{\mathrm{Inflex}}_{n,t}=w^{\mathrm{Inflex}0}_{n,t}
    \label{eq:inflexible_workload_balance}
\end{equation}
After spatial routing of interactive requests and temporal rescheduling of batch tasks, the total workload processed at DC \(n\) in period \(t\) is:
\begin{equation}
\begin{aligned}
    w_{n,t}&=w^{\mathrm{Inflex}}_{n,t}
    +w^{\mathrm I}_{n,t}+w^{\mathrm B}_{n,t}
\end{aligned}
    \label{eq:total_processed_workload}
\end{equation}
Equation~\eqref{eq:total_processed_workload} aggregates the three workload classes after scheduling. Equations~\eqref{eq:interactive_routing_balance} -- \eqref{eq:batch_annual_completion} ensure that workload scheduling can shift the location and time of power consumption without discarding the computing service. The processed workload \(w_{n,t}\), together with the number of active servers, determines server utilization through Eq.~\eqref{eq:server_utilization} and, consequently, the DC power demand seen by the power system. This establishes the connection between workload scheduling and the DC electricity demand entering the network model.

\section{Hosting capacity assessment methodology}
\label{sec:methodology}

\subsection{Problem statement}
\label{sec:problem_statement}

\begin{figure*}[t]
    \centering
    \includegraphics[width=\textwidth]{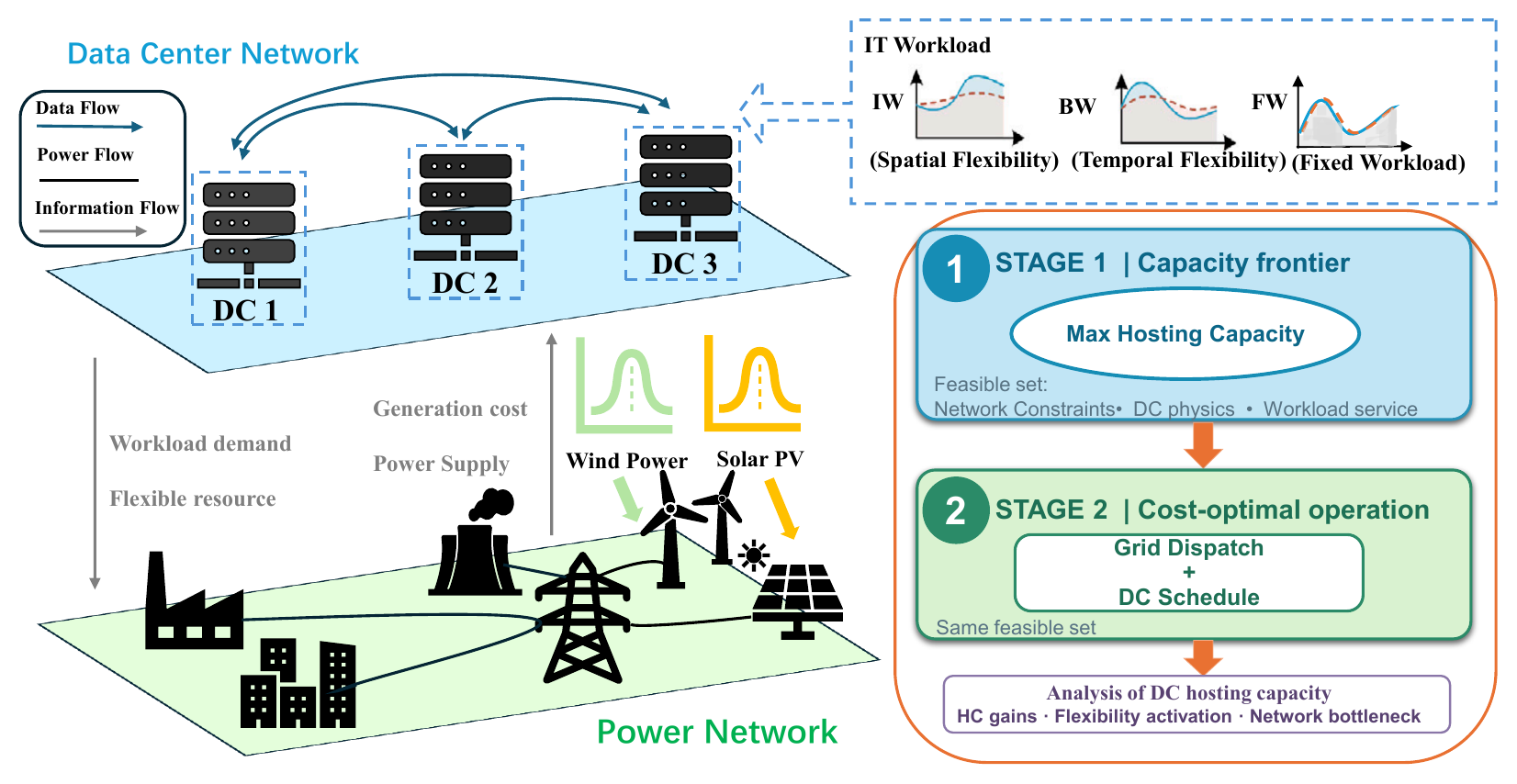}
    \caption{Illustration of the proposed hosting capacity assessment of data centers.}
    \label{fig:proposed_hc_assessment}
\end{figure*}

The mismatch between DC development and grid expansion timelines means that the network may not expand quickly enough to accommodate all proposed DC connections. Harvesting the flexibility within DCs therefore offers a potential solution for accommodating additional DC capacity through more effective use of the existing network capacity. Against this background, the problem considered in this paper is to quantify the maximum DC hosting capacity that can be connected to the transmission network. Each MW of admitted DC capacity is associated with a corresponding volume of computational task arrivals. Feasible hosting capacity must therefore support the complete computing service obligation over the annual horizon. The resulting hosting capacity represents the maximum compute-backed DC hosting capacity that can be accommodated by the network \cite{FangLi2026DCHostingCapacity}.

To achieve electricity-computation co-optimization, the problem is formulated from a coordinated system planning perspective. Fig.~\ref{fig:proposed_hc_assessment} illustrates the coupling between DC workload management and power system operation. On the DC side, interactive workloads may be migrated among geographically distributed DCs, providing spatial flexibility, while batch workloads may be deferred within service windows, providing temporal flexibility. The scheduled workload at each DC determines IT power consumption and, consequently, total DC power demand. On the power system side, DC power demand enters the nodal power balance and affects generation, voltage angles, and transmission flows.

\subsection{Proposed two-stage framework}
\label{sec:two_stage_framework}

\subsubsection{Framework overview}
\label{sec:framework_overview}

The proposed framework adopts a sequential lexicographic structure that assigns distinct planning functions to hosting capacity assessment and system operational optimization. The two stages therefore follow the sequence of first determining how much DC hosting capacity can be connected and where it can be accommodated, and then determining how the resulting DC portfolio and power system should be operated.

As illustrated in Fig.~\ref{fig:proposed_hc_assessment}, the DC hosting capacity vector \(\boldsymbol H_c^\star\) provides the coupling between the two stages. This two-stage structure provides complementary planning and operational outputs. Differences in the Stage 1 frontiers across flexibility configurations quantify the individual and coordinated hosting capacity values of spatial and temporal workload flexibility. The corresponding Stage 2 solutions reveal how flexibility is deployed across DCs and operating hours, and how the resulting DC power demand redistribution influences generation and transmission network conditions. The framework therefore connects the hosting capacity value of DC flexibility with the operational mechanisms.

\subsubsection{\protect\raggedright Stage 1: hosting capacity assessment}
\label{sec:stage1_hc}

Stage 1 determines the DC capacities and workload schedules and is formulated as follows:
\begin{equation}
    H_c^{\star}=
    \max_{\mathcal V_c^{1}}
    \sum_{n\in\mathcal D}H_{n,c}
    \label{eq:stage1_hosting_capacity}
\end{equation}
 The decision set includes the site capacities, workload scheduling and DC operating variables, and network operating variables:
\begin{equation}
\begin{aligned}
\mathcal V_c^{1}=\{&H_{n,c},w^{\mathrm{Inflex}}_{n,t},w^{\mathrm I}_{n,t},
w^{\mathrm B}_{n,t},w_{n,t},
w^{\mathrm I}_{n,n',t},w^{\mathrm B}_{n,t,t'},\\
&m_{n,t},u_{n,t},p^{\mathrm{IT}}_{n,t},p^{\mathrm{DC}}_{n,t},
p^{\mathrm G}_{g,t},p^{\mathrm R}_{r,t},\theta_{b,t},f_{\ell,t}\}
\end{aligned}
    \label{eq:stage1_decision_variables}
\end{equation}

\noindent\textit{subject to:}
\begin{align*}
\text{DC power and IT constraints:}\quad
    &\eqref{eq:dc_power_pue}\text{--}\eqref{eq:server_utilization},\\
\text{Workload constraints:}\quad
    &\eqref{eq:interactive_routing_balance}\text{--}
      \eqref{eq:total_processed_workload},\\
\end{align*}

Each DC \(n\) is connected at transmission network bus \(i(n)\). Its admitted hosting capacity \(H_{n,c}\) bounds the facility power that can be drawn at this connection point. Accordingly, the scheduled DC power demand must satisfy:
\begin{equation}
    H_{n,c}\geq0,\qquad
    0\leq p^{\mathrm{DC}}_{n,t}\leq H_{n,c}
    \label{eq:dc_connection_limit}
 \end{equation}
The admitted capacity is a planning decision, whereas \(p^{\mathrm{DC}}_{n,t}\) varies across operating periods with the workload schedule and server operation. Feasibility requires the connection limit to hold in every period.

Conventional and renewable generation supply both the existing system load and the additional DC demand. Let \(p^{\mathrm G}_{g,t}\) denote conventional generation, and
\(p^{\mathrm R}_{r,t}\) denote actual renewable generation. Generator operation
requires:
\begin{equation}
    \underline p^{\mathrm G}_g\leq p^{\mathrm G}_{g,t}
    \leq\bar p^{\mathrm G}_g
    \label{eq:conventional_generation_limits}
\end{equation}
\begin{equation}
    0\leq p^{\mathrm R}_{r,t}\leq\bar p^{\mathrm R}_{r,t}
    \label{eq:renewable_generation_limits}
\end{equation}
Renewable curtailment is the portion of available renewable output that is not dispatched:
\begin{equation}
    p^{\mathrm{curt}}_{r,t}
    =\bar p^{\mathrm R}_{r,t}-p^{\mathrm R}_{r,t}\geq0
    \label{eq:renewable_curtailment}
\end{equation}

For transmission line \(\ell\), let \(o(\ell)\) and \(e(\ell)\) be its
sending and receiving buses, \(x_\ell\) denote its reactance, and
\(\bar f_\ell\) be its active power rating. Under the DC power flow
representation:
\begin{equation}
    f_{\ell,t}=\frac{S^{\mathrm{base}}}{x_\ell}
    \left(\theta_{o(\ell),t}-\theta_{e(\ell),t}\right)
    \label{eq:dc_line_flow}
\end{equation}
Network constraints are:
\begin{equation}
    -\bar f_\ell\leq f_{\ell,t}\leq\bar f_\ell,
    \qquad \ell\in\mathcal L,\;t\in\mathcal T,
    \label{eq:line_limits}
\end{equation}
\begin{equation}
    -\pi\leq\theta_{b,t}\leq\pi,
    \qquad b\in\mathcal N,\;t\in\mathcal T,
    \label{eq:angle_limits}
\end{equation}
\begin{equation}
    \theta_{b^{\mathrm{ref}},t}=0,
    \qquad t\in\mathcal T
    \label{eq:reference_angle}
\end{equation}
Let \(\mathcal G_b\), \(\mathcal R_b\), and \(\mathcal D_b\) collect the
generators, renewable units, and DCs at bus \(b\), and let
\(\mathcal L_b^{\mathrm{out}}\) and \(\mathcal L_b^{\mathrm{in}}\) collect
its oriented outgoing and incoming lines. Nodal balance is:
\begin{equation}
\begin{aligned}
    &\sum_{g\in\mathcal G_b}p^{\mathrm G}_{g,t}
    +\sum_{r\in\mathcal R_b}p^{\mathrm R}_{r,t}
    =p^{\mathrm{load}}_{b,t}+\sum_{n\in\mathcal D_b}p^{\mathrm{DC}}_{n,t}\\
    &\quad+\sum_{\ell\in\mathcal L_b^{\mathrm{out}}}f_{\ell,t}
    -\sum_{\ell\in\mathcal L_b^{\mathrm{in}}}f_{\ell,t},
    \qquad b\in\mathcal N,\;t\in\mathcal T
\end{aligned}
    \label{eq:nodal_power_balance}
\end{equation}

Together, these constraints require the computing service associated with each admitted capacity to be supported by feasible DC and power system operation over the annual horizon. The optimal value \(H_c^\star\) is therefore the aggregate hosting capacity frontier under the modeled workload, equipment, generation, and network constraints. The corresponding DC hosting capacity vector \(\boldsymbol H_c^\star\) is passed to Stage 2.

\subsubsection{\protect\raggedright Stage 2: Optimal operation and scheduling}
\label{sec:stage2_operation}

Stage 2 solves a DC optimal power flow problem to minimize the sum of generation and carbon costs and is formulated as follows:

\begin{equation}
    \min_{\mathcal V_c^{2}}\quad
    C^{\mathrm{oper}}_c,
    \qquad
    C^{\mathrm{oper}}_c=C^{\mathrm G}_c+C^{\mathrm{carbon}}_c
    \label{eq:stage2_system_cost}
\end{equation}

 Unlike Stage 1, the Stage 2 decision set excludes the site capacities, which are now fixed parameters:
\begin{equation}
\begin{aligned}
\mathcal V_c^{2}=\{&w^{\mathrm{Inflex}}_{n,t},w^{\mathrm I}_{n,t},
w^{\mathrm B}_{n,t},w_{n,t},
w^{\mathrm I}_{n,n',t},w^{\mathrm B}_{n,t,t'},\\
&m_{n,t},u_{n,t},p^{\mathrm{IT}}_{n,t},p^{\mathrm{DC}}_{n,t},
p^{\mathrm G}_{g,t},p^{\mathrm R}_{r,t},\theta_{b,t},f_{\ell,t}\}
\end{aligned}
    \label{eq:stage2_decision_variables}
\end{equation}
The conventional generation cost is:
\begin{equation}
    C^{\mathrm G}_c=\Delta t\sum_{t\in\mathcal T}\sum_{g\in\mathcal G}
    \bigl[c_{2,g}(p^{\mathrm G}_{g,t})^2+c_{1,g}p^{\mathrm G}_{g,t}+c_{0,g}\bigr]
    \label{eq:generation_cost}
\end{equation}
Conventional generation also incurs a carbon cost. With generator emission factor \(\gamma_g\), total emissions and the
corresponding carbon cost are
\begin{equation}
\begin{aligned}
    E^{\mathrm{carbon}}_c=\Delta t\sum_{t\in\mathcal T}
    \sum_{g\in\mathcal G}\gamma_gp^{\mathrm G}_{g,t},\\
    C^{\mathrm{carbon}}_c=c^{\mathrm{CO2}}E^{\mathrm{carbon}}_c
\end{aligned}
    \label{eq:emissions_and_carbon_cost}
\end{equation}

\noindent\textit{subject to:}
\begin{align*}
\text{DC and workload constraints:}\quad
    &\eqref{eq:dc_power_pue}\text{--}\eqref{eq:total_processed_workload},\\
\text{Transmission network constraints:}\quad
    &\eqref{eq:dc_connection_limit}\text{--}\eqref{eq:nodal_power_balance},
\end{align*}

The Stage 2 objective in Eq.~\eqref{eq:stage2_system_cost} optimizes system
operation at the hosting capacity frontier by minimizing total system operational cost.
It combines the quadratic conventional generation cost in
Eq.~\eqref{eq:generation_cost} with the explicit carbon cost in
Eq.~\eqref{eq:emissions_and_carbon_cost}, thereby considering the emissions associated with dispatch and favoring lower carbon supply~\cite{Andrianesis2021EmissionsCost}. The resulting solution describes cost-minimizing annual operation for the capacity allocation obtained in Stage 1.

\makeatletter
\let\compactoriginalsubsubsection\subsubsection
\renewcommand\subsubsection{\@startsection{subsubsection}{3}{\z@}%
  {12\p@ \@minus 3\p@}{\p@}{\normalfont\normalsize\itshape}}
\makeatother
\subsubsection{Compact two-stage formulation}
\let\subsubsection\compactoriginalsubsubsection
\label{sec:compact_two_stage_formulation}

Following the detailed stage formulations, the complete lexicographic
structure is written in the hierarchical form of
Eq.~\eqref{eq:compact_two_stage_formulation}. The framework first maximizes aggregate DC hosting capacity and then minimizes annual operating cost at the resulting capacity allocation. The two stages are summarized as

\begin{equation}
\begin{aligned}
\text{Stage 1:}\qquad
&H_c^{\star}
=\max_{\mathcal V_c^{1}}
\sum_{n\in\mathcal D}H_{n,c},\\[2mm]
\text{Stage 2:}\qquad
&\min_{\mathcal V_c^{2}}
\left(C_c^{\mathrm G}+C_c^{\mathrm{carbon}}\right),\\
&\text{with }H_{n,c}=H_{n,c}^{\star},
\qquad n\in\mathcal D
\end{aligned}
\label{eq:compact_two_stage_formulation}
\end{equation}

Both stages are subject to the DC power and IT constraints in Eqs.~\eqref{eq:dc_power_pue}--\eqref{eq:server_utilization}, the workload conservation and completion constraints in Eqs.~\eqref{eq:interactive_routing_balance}--\eqref{eq:total_processed_workload}, the DC connection constraint in Eq.~\eqref{eq:dc_connection_limit}, and the generation and transmission-network constraints in Eqs.~\eqref{eq:conventional_generation_limits}--\eqref{eq:nodal_power_balance}. The generation and carbon costs are defined in Eqs.~\eqref{eq:generation_cost} and~\eqref{eq:emissions_and_carbon_cost}, respectively.

This sequential procedure gives priority to hosting capacity maximization and then minimizes operating cost for the optimized DC hosting capacity.

\section{Case studies}
\label{sec:case_studies}

This section evaluates the proposed framework on a modified IEEE-118 transmission network. The models were implemented in Julia 1.12.5 and solved using Gurobi 13.0.0 \cite{Gurobi2025Manual}. All computations were executed on a
workstation equipped with a 14-core Intel Core i7 CPU @2.30 GHz and 16 GB RAM.

\subsection{Experiment settings}
\label{sec:experiment_settings}

\begin{figure*}[!t]
    \centering
    \includegraphics[width=\textwidth,height=0.359976\textheight,keepaspectratio]{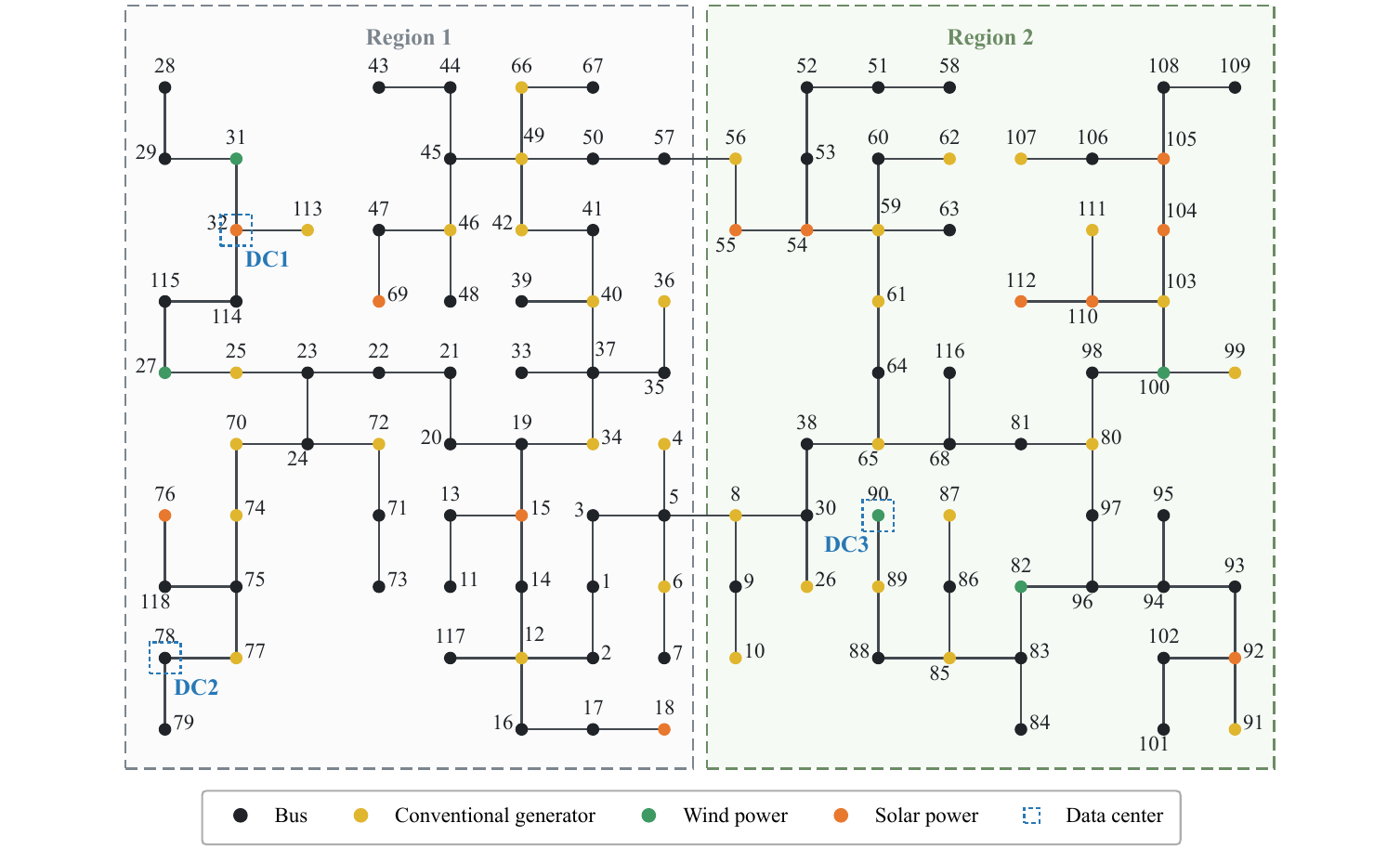}
    \caption{Topology of the modified IEEE-118 transmission system and locations of the three data centers.}
    \label{fig:case_study}
\end{figure*}

The study uses a modified IEEE-118 transmission system, whose source
provides annual network, generator, load, wind, and solar data
\cite{Pena2018NREL118}. Fig.~\ref{fig:case_study} summarizes the modified transmission network and the prospective DC locations. The system contains 118 buses and 186 lines, with DCs connected at buses 32, 78, and 90. All simulations are conducted over an 8760-hour year with a 1-hour resolution.

Workload arrival profiles are derived from the IN2P3 2024 Workload
Dataset \cite{CochardSimon2026IN2P3}. Based on \cite{Klingert2020SpinningGold} and characteristics of DC workloads, the shares of interactive, batch, and fixed workloads are assumed to be 40\%, 20\%, and 40\% respectively (Fig.~3). The DC configuration parameters are listed in Table 2, based on \cite{Tencent2018RDMA,Zhang2023InteractiveIntegration}.

\begin{figure}[!t]
    \centering
    \includegraphics[width=\columnwidth]{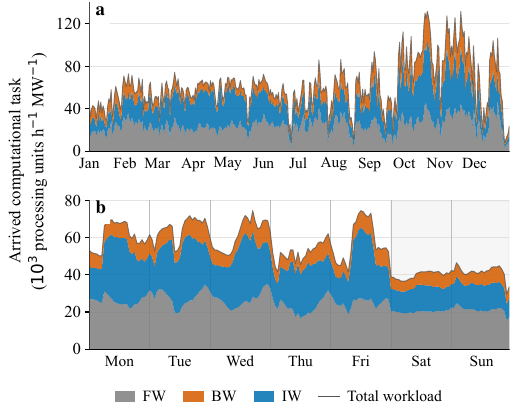}
    \caption{Annual (a) and weekly (b) computational task arrivals at DC1.}
    \label{fig:dc1_arrived_computational_task}
\end{figure}
\begin{table}[H]
\centering
\caption{Configuration of DCs.}
\label{tab:base_parameters}
\footnotesize
\begin{tabular}{@{}l@{\hspace{2em}}l@{}}
\toprule
Parameter & Value \\
\midrule
\(\eta^{\mathrm{DC}}\) & \(1.20~\) \\
\(p^{\mathrm{idle}}\) & \(1.3~(\mathrm{kW})\) \\
\(p^{\mathrm{peak}}\) & \(1.9~(\mathrm{kW})\) \\
\(\mu\) & \(500~(\text{workload units}/(\mathrm{h}))\) \\
\(\bar U\) & \(0.80~\) \\
\(T^{\mathrm{D}}\) & \(6~(\mathrm{h})\) \\
\bottomrule
\end{tabular}
\end{table}

To quantify the hosting capacity value of DC workload flexibility, the following four cases are simulated and compared:
\begin{itemize}
    \item \textbf{Case 1:} No workload flexibility (NF).
    \item \textbf{Case 2:} Temporal flexibility of batch workloads (TF).
    \item \textbf{Case 3:} Spatial flexibility of interactive workloads (SF).
    \item \textbf{Case 4:} Coordinated spatial and temporal workload flexibility (STF).
\end{itemize}

\subsection{Simulation results analysis}
\label{sec:simulation_results_analysis}
\begin{figure*}[t]
    \centering
    \includegraphics[width=\textwidth,height=0.35\textheight,keepaspectratio]{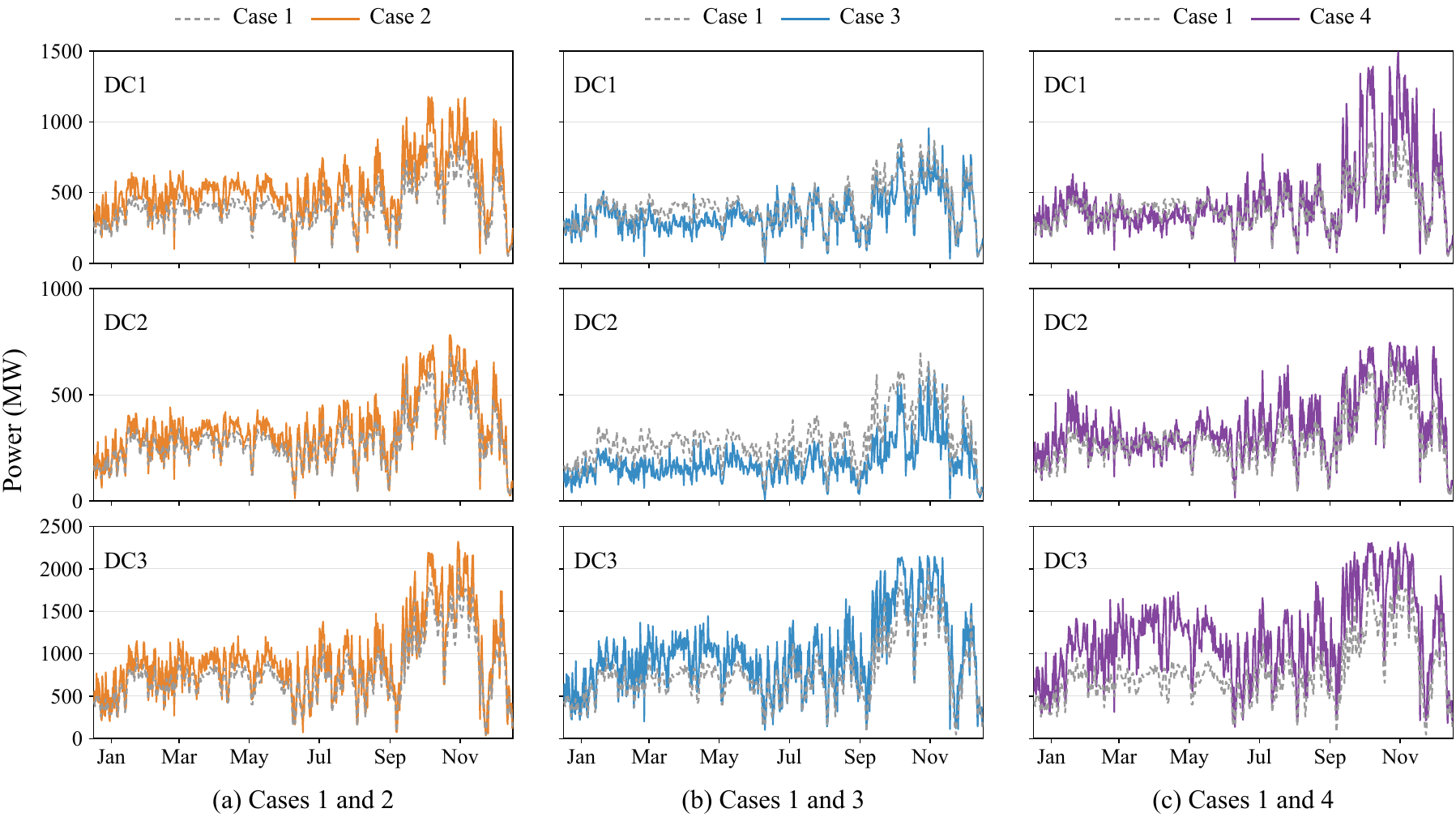}
    \caption{Power consumption of DCs under (a) Cases 1 and 2, (b) Cases 1 and 3, and (c) Cases 1 and 4.}
    \label{fig:dc_power_consumption_four_cases}
\end{figure*}

Fig.~\ref{fig:dc_power_consumption_four_cases} compares the annual DC power profiles under the four flexibility configurations. Case~2 broadly preserves the consumption pattern of each DC in Case~1, while reshaping the temporal profile as batch workloads can be deferred within their allowable window. Case~3 exhibits a distinct spatial redistribution pattern: DC1 and DC2 consume less power than case 1 over long periods, whereas DC3 accommodates a larger share of workload. Case~4 combines these patterns with particularly pronounced high-power periods at DC3. Importantly, the differences among the trajectories reflect both the larger workload scale enabled by the increased hosting capacity and the optimized spatial or temporal scheduling. The corresponding workload migration and deferral volumes are examined in Section~\ref{sec:analysis_of_flexibility_in_dcs}.

The following metrics quantify the hosting capacity gains relative to different cases. Case 1 (NF) serves as the benchmark for the other cases. The absolute and relative hosting capacity gains are defined as:
\begin{equation}
    V^{\mathrm{HC}}_c=H_c^{\star}-H_{\mathrm{NF}}^{\star}
    \label{eq:absolute_capacity_value}
\end{equation}

\begin{equation}
    v^{\mathrm{HC}}_c=\,
    \frac{H_c^{\star}-H_{\mathrm{NF}}^{\star}}
    {H_{\mathrm{NF}}^{\star}}
    \label{eq:relative_capacity_value}
\end{equation}
The four case comparison also quantifies the interaction between spatial and temporal flexibility:
\begin{equation}
    V_{\mathrm{interaction}}^{\mathrm{HC}}=
    H_{\mathrm{STF}}^{\star}-H_{\mathrm{SF}}^{\star}
    -H_{\mathrm{TF}}^{\star}+H_{\mathrm{NF}}^{\star}
    \label{eq:spatiotemporal_interaction}
\end{equation}

To relate the HC gain to the accompanying change in annual
system operational cost, the HC cost efficiency, expressed in MW per million pounds, is defined as:
\begin{equation}
    \mathrm{HCCE}_c=
    \frac{V_c^{\mathrm{HC}}}
    {\left(C_c^{\mathrm{oper}}-C_{\mathrm{NF}}^{\mathrm{oper}}\right)}
    \qquad
    \label{eq:hc_cost_efficiency}
\end{equation}

\begin{table*}[t]
\centering
\caption{Simulation metrics under four cases.}
\label{tab:simulation_metrics_four_cases}
\footnotesize
\setlength{\tabcolsep}{5pt}
\begin{tabular}{crrrrrrr}
\toprule
Case & \multicolumn{4}{c}{Hosting capacity} & System operating & $\mathrm{HCCE}_c$ & Solution time \\
\cmidrule(lr){2-5}
 & $H_c^{\star}$ (MW) & $V_c^{\mathrm{HC}}$ (MW) & $v_c^{\mathrm{HC}}$ (\%) & $V_{\mathrm{interaction}}^{\mathrm{HC}}$ (MW) & cost ($10^9\,\pounds$) & (MW/$10^6\,\pounds$) & (s) \\
\midrule
1 & 4635.1 & -- & -- & -- & 4.405 & -- & 1190.15 \\
2 & 5620.5 & 985.4 & 21.26 & -- & 4.546 & 6.99 & 1584.77 \\
3 & 4884.2 & 249.1 & 5.37 & -- & 4.442 & 6.84 & 2060.16 \\
4 & 6231.5 & 1596.4 & 34.44 & 361.9 & 4.641 & 6.77 & 3167.51 \\
\bottomrule
\end{tabular}
\end{table*}

Table~\ref{tab:simulation_metrics_four_cases} establishes Case~1 as the
inflexible benchmark, with an aggregated hosting capacity of 4635.1~MW.
Enabling temporal flexibility alone in Case~2 raises this frontier to
5620.5~MW, corresponding to a gain of 985.4~MW (21.26\% improvement over case 1). Spatial flexibility alone increases HC by 249.1 MW (5.37\%), raising the frontier to 4884.2 MW. The coordinated temporal and spatial flexibility in Case~4 achieves the largest hosting capacity, 6231.5~MW, which is 1596.4~MW (34.44\% above the benchmark). This result reflects the greater flexibility value of temporal deferral, which can shift workload away from network-constrained hours and exploit headroom available in subsequent periods, whereas spatial shifting is limited to the headroom available at alternative DC locations within the same hour. Coordinating the two mechanisms produces the largest HC gain in this case study.

Substituting the results reported in
Table~\ref{tab:simulation_metrics_four_cases} into
Eq.~\eqref{eq:spatiotemporal_interaction} gives
$V_{\mathrm{interaction}}^{\mathrm{HC}}=361.9$~MW, equivalent to 7.81\% of the Case~1 frontier. The
coordinated HC gain therefore exceeds the sum of the two individual gains by
361.9~MW. Equivalently, adding spatial flexibility to the temporally flexible case yields an additional 611.0~MW of HC, whereas adding temporal flexibility to the spatially flexible case yields a further 1347.3~MW. This positive interaction indicates that temporal and spatial flexibility are complementary in expanding the feasible hosting capacity region, rather than providing merely additive benefits.

\begin{figure}[t]
    \centering
    \includegraphics[width=\columnwidth]{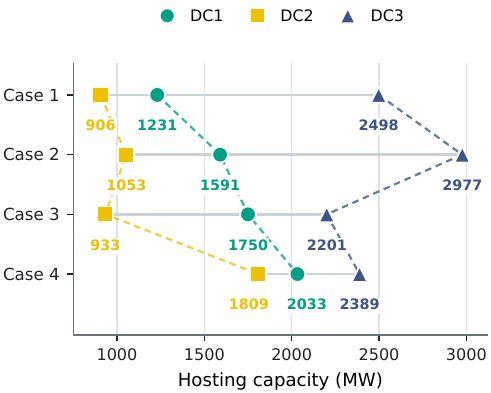}
    \caption{Hosting capacity of DCs under four cases.}
    \label{fig:dc_hosting_capacity_four_cases}
\end{figure}

Fig.~\ref{fig:dc_hosting_capacity_four_cases} compares the hosting capacity of the three DCs across the four cases. Under Case~1, DC3 receives the largest HC, approximately
2498~MW, compared with 1231~MW at DC1 and 906~MW at DC2. Temporal flexibility
in Case~2 increases the HC assigned to all three DCs. Spatial
flexibility in Case~3 instead reshapes the allocation: more HC is
supported at DC1 while the HC at DC3 is lower than in Case~1,
but the aggregate HC of the three DCs still increases. Under coordinated temporal and spatial flexibility, the HC at DC1 and DC2 rise to approximately 2033 and 1809~MW,
respectively, while DC3 retains approximately 2389~MW. Relative to Case~1,
the increases at DC1 and DC2 therefore more than compensate for the modest
reduction at DC3. This reallocation represents the optimization outcome through which flexibility redistributes the additional hosting capacity across DC locations.

The system operational costs reported in Table~\ref{tab:simulation_metrics_four_cases} should be interpreted separately for each case, as each value corresponds to the hosting capacity frontier of the respective case. Relative to Case~1, Cases~2--4 support more hosting capacity and have higher annual costs, reaching \pounds4.546, \pounds4.442, and \pounds4.641 billion, respectively. The corresponding HCCE values, calculated from the cost, remain within a range of 6.77--6.99~MW per million pounds. Case~4 thus delivers the largest absolute HC gain, although its incremental capacity-to-cost ratio is slightly lower than those of Cases 2 and 3. The richer flexibility formulations also increase computational complexity, resulting in longer solution times, from 1190.15~s in Case~1 to 3167.51~s in Case~4. These results quantify the HC frontier and its operational consequence. The associated changes in network operation are examined in Section~\ref{sec:analysis_of_system_impact}.

\subsection{Analysis of flexibility in DCs}
\label{sec:analysis_of_flexibility_in_dcs}

\begin{figure*}[t]
    \centering
    \includegraphics[width=\textwidth,height=0.45\textheight,keepaspectratio]{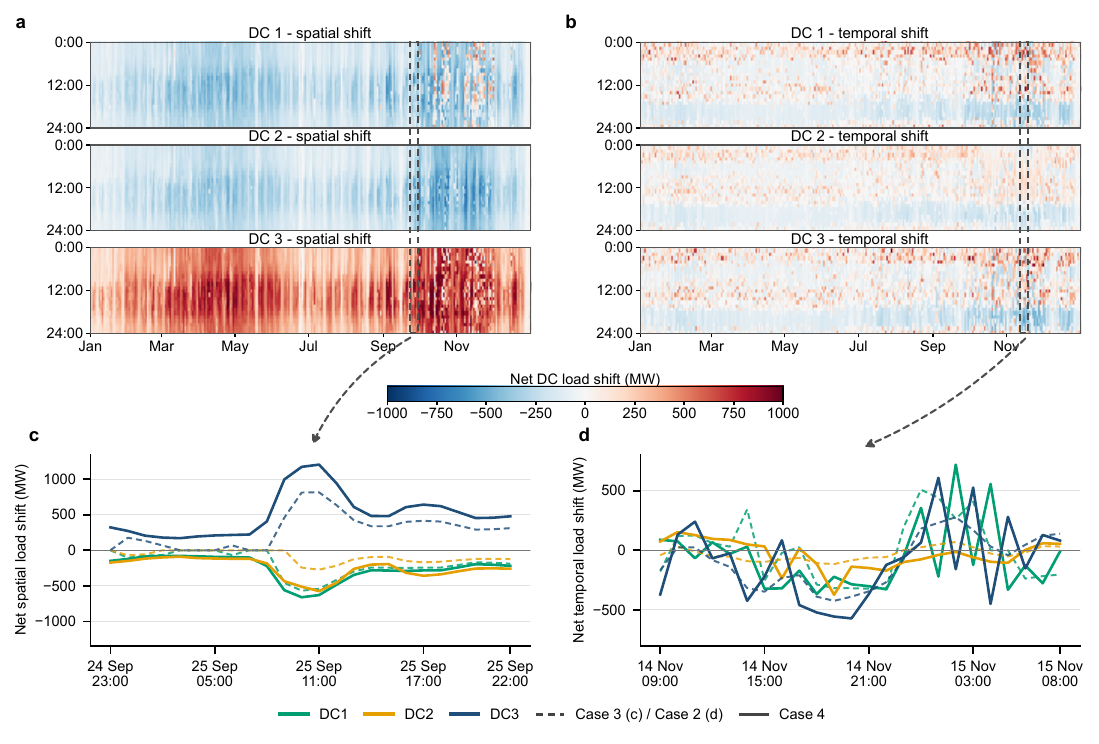}
    \caption{Annual patterns of DC load shifting. \textbf{a}, \textbf{b}\textbf{:} Spatial and temporal load shifts under Case~4. Color mapping represents the quantity of net load shifts from other DCs/other times (positive values) or shifts to other DCs/other times (negative values). \textbf{c}, \textbf{d}\textbf{:} Typical activation windows of spatial and temporal flexibility under different cases.     }
    \label{fig:dc_spatiotemporal_load_shift_case4}
\end{figure*}

The HC gains reported in Section~4.2 raise a further question:
how is the available DC flexibility actually utilized in the operational schedule? Figure~\ref{fig:dc_spatiotemporal_load_shift_case4} addresses this
question by showing the actual spatial and temporal workload shifting in Case~4. The widespread colored regions in Figs.~\ref{fig:dc_spatiotemporal_load_shift_case4}(a)
and (b) demonstrate that both spatial and temporal flexibility are activated throughout the year. The two maps therefore illustrate how spatial and temporal flexibility reshape the location and time of DC power demand in the optimized schedule.

The annual pattern of spatial flexibility in
Fig.~\ref{fig:dc_spatiotemporal_load_shift_case4}(a) has a clear directional
structure. DC1 and DC2 predominantly exhibit negative workload shifts, whereas DC3 experiences positive shifts, with the spatial redistribution becoming more pronounced in the later part of the year. The simultaneous occurrence of opposite signs across the three sites is consistent with workload migration: DC1 and DC2 primarily act as workload-sending DC sites, whereas DC3 predominantly serves as the workload-receiving DC site. The persistence of this pattern across the annual map shows that spatial flexibility is used as a recurring portfolio-level response, although its intensity varies with date and hour.

The annual pattern of temporal flexibility in
Fig.~\ref{fig:dc_spatiotemporal_load_shift_case4}(b) exhibits a distinct operational pattern. Each DC displays alternating positive and negative regions, reflecting the temporal redistribution of workload across the day. Overall, workload is mainly shifted away from the morning and evening periods and to midday and overnight hours. These shifts are unevenly distributed across the year, and the late-year maps show more pronounced variations, particularly at DC1 and DC3. Temporal flexibility therefore reshapes DC power demand at each site, rather than reproducing the redistribution across DCs observed for spatial flexibility.

The enlarged windows in
Figs.~\ref{fig:dc_spatiotemporal_load_shift_case4}(c) and (d) further highlight these distinct spatial and temporal operating patterns. In the September spatial window, the solid Case~4 profiles show an increase at DC3 accompanied by reductions at DC1 and DC2. The dashed SF profiles follow the same direction of shifting but with smaller excursions over the main activated period. In the November temporal window, the solid profiles alternate between negative and positive shifts at the three DCs, displaying the deferral and subsequent recovery of workload within the window. Relative to the dashed SF and TF profiles in panels (c) and (d), respectively, the STF case modifies both the magnitude and timing of the flexibility response at each DC, with the most pronounced differences observed at DC1 and DC3.

Taken together, the annual maps and activation windows illustrate how the two flexibility mechanisms complement each other at the DC level. Spatial flexibility determines where eligible workloads are processed, whereas temporal flexibility determines when they are processed. Their coordinated utilization therefore provides two distinct dimensions of adjustment in the spatio-temporal scheduling of DC power demand. The resulting impacts on the power system are examined separately in the next section.

\subsection{Analysis of system impact}
\label{sec:analysis_of_system_impact}

Sections 4.2 and 4.3 quantify the hosting capacity gains and characterize the spatial and temporal workload shifting. This section examines how workload shifting reshapes transmission line loading and allows additional DC hosting capacity to be accommodated within the existing network.

\begin{figure*}[t]
    \centering
    \includegraphics[width=0.8  \textwidth,keepaspectratio]{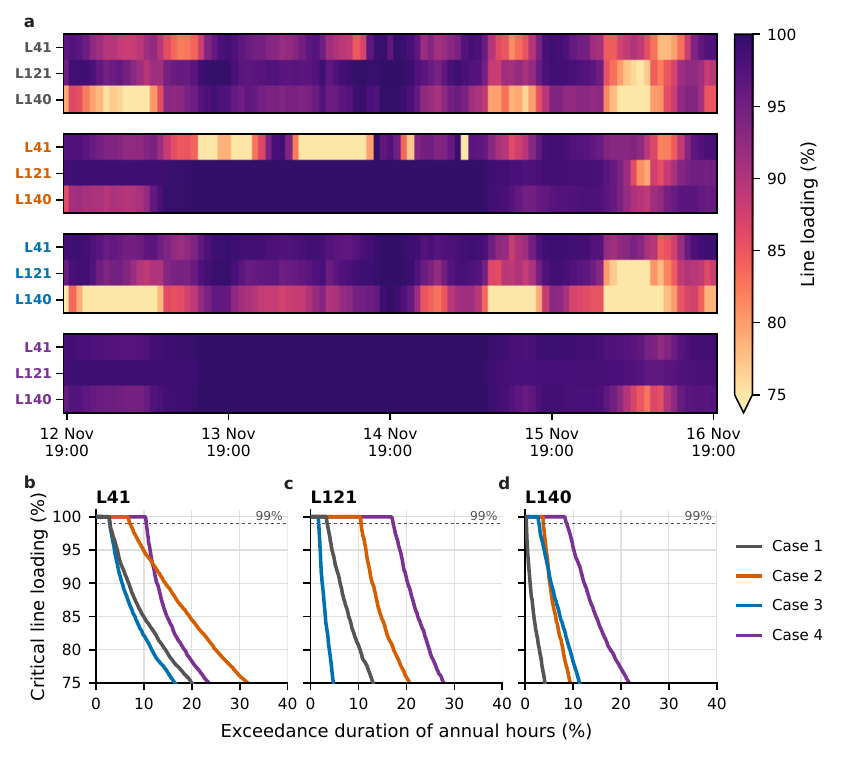}
    \caption{System impacts of data center flexibility under the four cases. \textbf{a:} Loading of critical lines (L41, L121, and L140) during the peak period. \textbf{b, c, d:} Annual loading duration curves for L41, L121, and L140 under the four cases.}
    \label{fig:system_impact}
\end{figure*}

Fig.~\ref{fig:system_impact}(a) illustrates the loading patterns of L41, L121, and L140 during the selected peak period. Under Case~1 (NF), high loading is particularly evident on L121 and L140 around the middle of the period, while L41 exhibits more pronounced temporal fluctuations. Under Case 2 (TF), near-limit loading on L121 and L140 persists over a broader sequence of hours, reflecting how the larger volume of workload is accommodated by scheduling its processing across time. Case 3 (SF) produces a distinct spatial response: loading on L41 becomes more sustained, whereas L140 experiences longer intervals of relatively low loading. This redistribution demonstrates how shifting workloads across DC locations reshapes transmission line loading and alters the network headroom to accommodate additional DC demand. Under Case 4 (STF), high line loading extends to all three lines over a larger portion of the period. Therefore, by scheduling DC power demand across time and location, flexibility reshapes transmission line loading and enables existing network headroom to be utilized more effectively.

The annual loading duration curves in Figs.~\ref{fig:system_impact}(b)–(d) show how these operational differences extend over the year. Case 4 (STF) exhibits the longest duration of near-limit operation on all three transmission lines, indicating more sustained utilization of their existing transmission capacity at the increased DC hosting capacity. The individual flexibility cases reveal distinct responses. Relative to Case~1 (NF), Case 3 (SF)  shortens near-limit operation on L121 and extends it on L140, demonstrating the redistribution of annual transmission line loading associated with spatial flexibility. Case 2 (TF) extends near-limit operation on all three lines through temporal flexibility. On L41, Case 2 (TF) also maintains elevated loading for longer at lower loading thresholds. Temporal and spatial flexibility therefore produce distinct annual loading patterns on the same transmission line over the year.

Together, these results explain the network mechanism underlying the hosting capacity gains. Spatial flexibility changes the distribution of demand among DC locations, while temporal flexibility changes its distribution across operating hours. Their coordinated use provides additional ways to supply the growing computational workload within the available transmission capacity. The more sustained loading observed under STF accompanies the increase in hosting capacity from 4635.1 MW under NF to 6231.5 MW, corresponding to a 34.44\%  improvement reported in Table 3. DC flexibility therefore increases the usable network headroom for additional DC connection through coordinated spatial and temporal scheduling of DC power demand.

\section{Conclusions and future work}
\label{sec:conclusions_and_future_work}

This paper developed a two-stage electricity--computation co-optimization framework for quantifying the spatial and temporal DC workload flexibility value in hosting capacity. The framework integrates workload scheduling, DC power consumption, generation dispatch, and transmission constraints over a complete 8760-hour year. The proposed two-stage framework first identifies the maximum aggregate DC hosting capacity in Stage 1 and then determines the annual system operation and DC workload schedule at the resulting hosting capacity allocation in Stage 2. This formulation establishes a direct connection between the planning value of DC flexibility and the operational decisions through which that value is realized.

The modified IEEE-118 case study demonstrates that both temporal and spatial workload flexibility increase the DC hosting capacity of the transmission network. Relative to the inflexible case, temporal and spatial flexibility provide gains of 985.4~MW (21.26\%) and 249.1~MW (5.37\%), respectively, while their coordinated use increases the hosting capacity to 6231.5~MW, representing a gain of 1596.4~MW (34.44\%). The coordinated gain exceeds the sum of the stand-alone gains by 361.9~MW (7.81\%), demonstrating the complementary value of coordinated flexibility across locations and time. This value is realized through distinct operational mechanisms: spatial flexibility redirects interactive workloads primarily from DC1 and DC2 to DC3, whereas temporal flexibility defers batch workloads from morning and evening periods to available midday and overnight hours. Their coordinated activation consequently reshapes both the hosting capacity allocation and the DC power demand profile. The resulting redistribution of line loading among L41, L121, and L140, together with the longer duration of near-limit operation at the optimized hosting capacity, indicates fuller utilization of the existing transmission network capacity. Workload flexibility therefore enables the network to accommodate additional DC hosting capacity by coordinating where and when computational workloads are served.

Future work can extend the framework towards uncertainty-aware connection assessment and its operational implementation. Stochastic and distributionally robust optimization can incorporate variations in computational task arrivals, conventional system demand, and renewable availability to obtain security and reliability-qualified hosting capacity values. Matched capacity comparisons at a common DC hosting capacity can quantify how workload flexibility affects its system impact in addition to its hosting capacity value. Finally, the optimized workload schedules can support the design of flexible or non-firm connection arrangements by defining activation conditions, workload service requirements, measurement and verification procedures, and the settlement mechanisms. These extensions would translate the quantified hosting capacity value of DC workload flexibility into implementable planning and operational solutions for power systems accommodating rapidly growing AI-era DC demand.

\FloatBarrier

\bibliographystyle{elsarticle-num}
\bibliography{references}

@techreport{IEA2025EnergyAI,
  author      = {{International Energy Agency}},
  title       = {Energy and {AI}},
  institution = {International Energy Agency},
  address     = {Paris},
  year        = {2025},
  month       = apr,
  url         = {https://www.iea.org/reports/energy-and-ai},
  urldate     = {2026-07-27}
}

@techreport{Shehabi2024DataCenterEnergy,
  author      = {Shehabi, Arman and Smith, Sarah J. and Hubbard, Alex and
                 Newkirk, Alexander and Lei, Nuoa and Siddik, Md AbuBakar and
                 Holecek, Billie and Koomey, Jonathan G. and Masanet, Eric R.
                 and Sartor, Dale A.},
  title       = {2024 United States Data Center Energy Usage Report},
  institution = {Lawrence Berkeley National Laboratory},
  address     = {Berkeley, CA},
  number      = {LBNL-2001637},
  year        = {2024},
  month       = dec,
  doi         = {10.71468/P1WC7Q},
  url         = {https://doi.org/10.71468/P1WC7Q}
}

@article{Masanet2020Recalibrating,
  author  = {Masanet, Eric and Shehabi, Arman and Lei, Nuoa and Smith, Sarah and
             Koomey, Jonathan},
  title   = {Recalibrating global data center energy-use estimates},
  journal = {Science},
  year    = {2020},
  volume  = {367},
  number  = {6481},
  pages   = {984--986},
  doi     = {10.1126/science.aba3758}
}

@article{Huang2020DataCenterProsumers,
  author  = {Huang, Pei and Copertaro, Benedetta and Zhang, Xingxing and
             Shen, Jingchun and L{\"o}fgren, Isabelle and R{\"o}nnelid, Mats and
             Fahlen, Jan and Andersson, Dan and Svanfeldt, Mikael},
  title   = {A review of data centers as prosumers in district energy systems:
             Renewable energy integration and waste heat reuse for district
             heating},
  journal = {Applied Energy},
  year    = {2020},
  volume  = {258},
  pages   = {114109},
  doi     = {10.1016/j.apenergy.2019.114109}
}

@article{Kim2017DispatchableLoads,
  author  = {Kim, Kibaek and Yang, Fan and Zavala, Victor M. and Chien, Andrew A.},
  title   = {Data Centers as Dispatchable Loads to Harness Stranded Power},
  journal = {IEEE Transactions on Sustainable Energy},
  year    = {2017},
  volume  = {8},
  number  = {1},
  pages   = {208--218},
  doi     = {10.1109/TSTE.2016.2593607}
}

@article{Radovanovic2023CarbonAware,
  author  = {Radovanovi{\'c}, Ana and Koningstein, Ross and Schneider, Ian and
             Chen, Bokan and Duarte, Alexandre and Roy, Binz and Xiao, Diyue and
             Haridasan, Maya and Hung, Patrick and Care, Nick and Talukdar,
             Saurav and Mullen, Eric and Smith, Kendal and Cottman, MariEllen
             and Cirne, Walfredo},
  title   = {Carbon-Aware Computing for Datacenters},
  journal = {IEEE Transactions on Power Systems},
  year    = {2023},
  volume  = {38},
  number  = {2},
  pages   = {1270--1280},
  doi     = {10.1109/TPWRS.2022.3173250}
}

@article{Andrianesis2021EmissionsCost,
  author  = {Andrianesis, Panagiotis and Biskas, Pandelis and
             Liberopoulos, George},
  title   = {Evaluating the cost of emissions in a pool-based electricity
             market},
  journal = {Applied Energy},
  year    = {2021},
  volume  = {298},
  pages   = {117253},
  doi     = {10.1016/j.apenergy.2021.117253}
}

@article{Kim2026FlexibilityAware,
  author  = {Kim, Dongjoo and Dong, Lin and Xie, Le},
  title   = {Flexibility-aware framework for efficient planner-initiated
             siting of data center},
  journal = {Nature Communications},
  year    = {2026},
  volume  = {17},
  number  = {1},
  pages   = {6512},
  doi     = {10.1038/s41467-026-72324-9}
}

@article{ZhangZavala2022Remunerating,
  author  = {Zhang, Weiqi and Zavala, Victor M.},
  title   = {Remunerating space--time, load-shifting flexibility from data
             centers in electricity markets},
  journal = {Applied Energy},
  year    = {2022},
  volume  = {326},
  pages   = {119930},
  doi     = {10.1016/j.apenergy.2022.119930}
}

@article{Ye2026Harvesting,
  author  = {Ye, Yujian and Ma, Ding and Wu, Yizhi and Hu, Heng and Zhang, Xi
             and Liu, Chuan and Xu, Dezhi},
  title   = {Harvesting spatial-temporal load migration flexibility of data
             centers: A chance-constrained bi-level optimization model with
             endogenously formed risk-reflective locational prices},
  journal = {Applied Energy},
  year    = {2026},
  volume  = {402},
  pages   = {126971},
  doi     = {10.1016/j.apenergy.2025.126971}
}

@article{Chen2021OperationalFlexibility,
  author  = {Chen, Sirui and Li, Peng and Ji, Haoran and Yu, Hao and Yan, Jinyue
             and Wu, Jianzhong and Wang, Chengshan},
  title   = {Operational flexibility of active distribution networks with the
             potential from data centers},
  journal = {Applied Energy},
  year    = {2021},
  volume  = {293},
  pages   = {116935},
  doi     = {10.1016/j.apenergy.2021.116935}
}

@article{Li2024ComputationPower,
  author  = {Li, Chuyi and Zheng, Kedi and Guo, Hongye and Kang, Chongqing and
             Chen, Qixin},
  title   = {Computation-Power Coupled Modeling for {IDCs} and Collaborative
             Optimization in {ADNs}},
  journal = {IEEE Transactions on Smart Grid},
  year    = {2024},
  volume  = {15},
  number  = {3},
  pages   = {2762--2775},
  doi     = {10.1109/TSG.2023.3321376}
}

@article{Cao2022DataDrivenFlexibility,
  author  = {Cao, Yujie and Cheng, Ming and Zhang, Sufang and Mao, Hongju and
             Wang, Peng and Li, Chao and Feng, Yihui and Ding, Zhaohao},
  title   = {Data-driven flexibility assessment for internet data center
             towards periodic batch workloads},
  journal = {Applied Energy},
  year    = {2022},
  volume  = {324},
  pages   = {119665},
  doi     = {10.1016/j.apenergy.2022.119665}
}

@article{Zeng2024BilevelPlanning,
  author  = {Zeng, Bo and Zhou, Yinyu and Xu, Xinzhu and Cai, Danting},
  title   = {Bi-level planning approach for incorporating the demand-side
             flexibility of cloud data centers under electricity-carbon
             markets},
  journal = {Applied Energy},
  year    = {2024},
  volume  = {357},
  pages   = {122406},
  doi     = {10.1016/j.apenergy.2023.122406}
}

@article{Shao2025StochasticPlanning,
  author  = {Shao, Zhentong and Yu, Nanpeng and Wong, Daniel},
  title   = {Stochastic long-term joint decarbonization planning for power
             systems and data centers: A case study in {PJM}},
  journal = {International Journal of Electrical Power \& Energy Systems},
  year    = {2025},
  volume  = {173},
  pages   = {111377},
  doi     = {10.1016/j.ijepes.2025.111377}
}

@article{Yang2023CarbonManagement,
  author  = {Yang, Ting and Jiang, Han and Hou, Yucheng and Geng, Yinan},
  title   = {Carbon Management of Multi-Datacenter Based On Spatio-Temporal
             Task Migration},
  journal = {IEEE Transactions on Cloud Computing},
  year    = {2023},
  volume  = {11},
  number  = {1},
  pages   = {1078--1090},
  doi     = {10.1109/TCC.2021.3130644}
}

@article{Wan2023NetworkTransition,
  author  = {Wan, Tong and Tao, Yuechuan and Qiu, Jing and Lai, Shuying},
  title   = {Internet data centers participating in electricity network
             transition considering carbon-oriented demand response},
  journal = {Applied Energy},
  year    = {2023},
  volume  = {329},
  pages   = {120305},
  doi     = {10.1016/j.apenergy.2022.120305}
}

@article{Cao2024NonWireAlternative,
  author  = {Cao, Yujie and Cao, Fang and Wang, Yajing and Wang, Jianxiao and
             Wu, Lei and Ding, Zhaohao},
  title   = {Managing data center cluster as non-wire alternative: A case in
             balancing market},
  journal = {Applied Energy},
  year    = {2024},
  volume  = {360},
  pages   = {122769},
  doi     = {10.1016/j.apenergy.2024.122769}
}

@article{Jian2024SupplyRestoration,
  author  = {Jian, Jie and Zhao, Jinli and Ji, Haoran and Bai, Linquan and
             Xu, Jing and Li, Peng and Wu, Jianzhong and Wang, Chengshan},
  title   = {Supply Restoration of Data Centers in Flexible Distribution
             Networks With Spatial-Temporal Regulation},
  journal = {IEEE Transactions on Smart Grid},
  year    = {2024},
  volume  = {15},
  number  = {1},
  pages   = {340--354},
  doi     = {10.1109/TSG.2023.3286844}
}

@article{Zhang2025ProactiveWorkload,
  author  = {Zhang, Yuanshi and Zou, Bokang and Jin, Xu and Luo, Yifu and
             Song, Meng and Ye, Yujian and Hu, Qinran and Chen, Qirui and
             Zambroni, Antonio Carlos},
  title   = {Mitigating power grid impact from proactive data center workload
             shifts: A coordinated scheduling strategy integrating synergistic
             traffic--data--power networks},
  journal = {Applied Energy},
  year    = {2025},
  volume  = {377},
  pages   = {124697},
  doi     = {10.1016/j.apenergy.2024.124697}
}

@article{Manassero2026ResilienceAware,
  author  = {Manassero, Jr., Giovanni and Rozenblit Tiferes, Rodrigo and
             Nascimento Ribeiro, Thamiris and Mazza, Andrea and Pons, Enrico},
  title   = {A resilience-aware methodology to support data center optimal
             allocation considering the transmission grid's hosting capacity},
  journal = {Applied Energy},
  year    = {2026},
  volume  = {423},
  pages   = {128351},
  doi     = {10.1016/j.apenergy.2026.128351}
}

@article{Samir2026AIHostingPEM,
  author  = {Samir, Omar and Zeineldin, Hatem and El-Saadany, Ehab F.},
  title   = {Hosting capacity of {AI} training data centers with {PEM}
             support: A two-level optimization approach},
  journal = {Applied Energy},
  year    = {2026},
  volume  = {421},
  pages   = {128231},
  doi     = {10.1016/j.apenergy.2026.128231}
}

@misc{FangLi2026DCHostingCapacity,
  author        = {Fang, Linhan and Li, Xingpeng},
  title         = {Hosting Capacity Assessment and Enhancement for Edge Data
                   Centers in Active Distribution Networks},
  year          = {2026},
  eprint        = {2606.01407},
  archiveprefix = {arXiv},
  primaryclass  = {eess.SY},
  doi           = {10.48550/arXiv.2606.01407},
  note          = {Accepted for presentation at the 2026 IEEE Industry
                   Applications Society Annual Meeting}
}

@article{Ding2022CoordinatedOperation,
  author  = {Ding, Zhaohao and Cao, Yujie and Zhang, Sufang and Wang, Peng and
             Liu, Jizhen and Cheng, Ming and Mao, Hongju},
  title   = {Coordinated Operation for Data Center and Power System in the
             Context of Energy Internet ({I}): Energy Demand Management Model
             of Data Center},
  journal = {Proceedings of the CSEE},
  year    = {2022},
  volume  = {42},
  number  = {9},
  pages   = {3161--3177},
  doi     = {10.13334/j.0258-8013.pcsee.210813}
}

@article{Pena2018NREL118,
  author  = {Pe{\~n}a, Ivonne and Martinez-Anido, Carlo Brancucci and
             Hodge, Bri-Mathias},
  title   = {An Extended {IEEE} 118-Bus Test System With High Renewable
             Penetration},
  journal = {IEEE Transactions on Power Systems},
  year    = {2018},
  volume  = {33},
  number  = {1},
  pages   = {281--289},
  doi     = {10.1109/TPWRS.2017.2695963}
}

@misc{CochardSimon2026IN2P3,
  author    = {Cochard, Guillaume and Simon, Bertrand},
  title     = {{CC-IN2P3} 2024 Workload Dataset},
  year      = {2026},
  publisher = {Zenodo},
  doi       = {10.5281/zenodo.18668107},
  url       = {https://doi.org/10.5281/zenodo.18668107}
}

@article{Klingert2020SpinningGold,
  author  = {Klingert, Sonja and Szilvas, Sebastian},
  title   = {Spinning gold from straw - evaluating the flexibility of data
             centres on power markets},
  journal = {Energy Informatics},
  year    = {2020},
  volume  = {3},
  pages   = {7},
  doi     = {10.1186/s42162-020-00110-y}
}

@misc{Tencent2018RDMA,
  author = {{Tencent Cloud Developer}},
  title  = {The first global-scale commercial deployment of {RDMA},
            {Tencent Cloud}'s core technology in computing},
  year   = {2018},
  url    = {https://cloud.tencent.com/developer/news/1261026}
}

@article{Zhang2023InteractiveIntegration,
  author  = {Zhang, Yuying and Zeng, Bo and Zhou, Yinyu and Xu, Hao and
             Liu, Wenxia},
  title   = {Research on interactive integration planning of data centers
             and distribution network driven by carbon emission reduction},
  journal = {Transactions of China Electrotechnical Society},
  year    = {2023},
  volume  = {38},
  number  = {23},
  pages   = {6433--6450},
  doi     = {10.19595/j.cnki.1000-6753.tces.221646}
}

@manual{Gurobi2025Manual,
  author       = {{Gurobi Optimization, LLC}},
  title        = {Gurobi Optimizer Reference Manual},
  year         = {2025},
  organization = {Gurobi Optimization, LLC},
  url          = {https://www.gurobi.com}
}

@techreport{VanGeet2024BestPractices,
  author      = {Van Geet, Otto and Sickinger, David},
  title       = {Best Practices Guide for Energy-Efficient Data Center Design},
  institution = {U.S. Department of Energy, Federal Energy Management Program},
  year        = {2024},
  month       = jul,
  url         = {https://www.energy.gov/sites/default/files/2024-07/best-practice-guide-data-center-design.pdf}
}

@techreport{EPRI2024PoweringDataCenters,
  author      = {{Electric Power Research Institute}},
  title       = {Powering Data Centers: {U.S.} Energy System and Emissions
                 Impacts of Growing Loads},
  institution = {Electric Power Research Institute},
  number      = {3002031198},
  year        = {2024},
  month       = oct,
  url         = {https://restservice.epri.com/publicdownload/000000003002031198/0/Product},
  urldate     = {2026-09-03}
}

@techreport{NERC2025EmergingLargeLoads,
  author      = {{North American Electric Reliability Corporation}},
  title       = {Characteristics and Risks of Emerging Large Loads},
  institution = {North American Electric Reliability Corporation},
  year        = {2025},
  month       = jul,
  url         = {https://www.nerc.com/globalassets/who-we-are/standing-committees/rstc/whitepaper-characteristics-and-risks-of-emerging-large-loads.pdf},
  urldate     = {2026-09-03}
}

@techreport{Ember2025GridsDataCentres,
  author      = {Cremona, Elisabeth and Czyzak, Pawel},
  title       = {Grids for Data Centres: Ambitious Grid Planning Can Win
                 Europe's {AI} Race},
  institution = {Ember},
  year        = {2025},
  month       = jun,
  url         = {https://ember-energy.org/app/uploads/2025/06/Report-Grids-for-data-centres-in-Europe-PDF.pdf},
  urldate     = {2026-09-03}
}

@article{Zhang2022HPCDemandResponse,
  author  = {Zhang, Yijia and Wilson, Daniel C. and Paschalidis, Ioannis Ch. and Coskun, Ayse K.},
  title   = {{HPC} Data Center Participation in Demand Response: An Adaptive Policy with {QoS} Assurance},
  journal = {IEEE Transactions on Sustainable Computing},
  year    = {2022},
  volume  = {7},
  number  = {1},
  pages   = {157--171},
  doi     = {10.1109/TSUSC.2021.3077254}
}

@article{Wu2023SpatiotemporalFlexibility,
  author  = {Wu, Zhaoyuan and Chen, Lin and Wang, Jianxiao and Zhou, Ming and Li, Gengyin and Xia, Qing},
  title   = {Incentivizing the Spatiotemporal Flexibility of Data Centers Toward Power System Coordination},
  journal = {IEEE Transactions on Network Science and Engineering},
  year    = {2023},
  volume  = {10},
  pages   = {1766--1778},
  doi     = {10.1109/TNSE.2023.3234445}
}

@inproceedings{Sukprasert2024Limitations,
  author    = {Sukprasert, Thanathorn and Souza, Abel and Bashir, Noman and Irwin, David and Shenoy, Prashant},
  title     = {On the Limitations of Carbon-Aware Temporal and Spatial Workload Shifting in the Cloud},
  booktitle = {Proceedings of the Nineteenth European Conference on Computer Systems},
  year      = {2024},
  pages     = {924--941},
  publisher = {Association for Computing Machinery},
  doi       = {10.1145/3627703.3650079}
}

@article{Riepin2025CleanComputing,
  author  = {Riepin, Iegor and Brown, Tom and Zavala, Victor M.},
  title   = {Spatio-temporal Load Shifting for Truly Clean Computing},
  journal = {Advances in Applied Energy},
  year    = {2025},
  volume  = {17},
  pages   = {100202},
  doi     = {10.1016/j.adapen.2024.100202}
}

\end{document}